# THE ROLE OF MAGNETIC FLUX EXPULSION TO REACH $Q_0>3\times10^{10}$ IN SRF CRYOMODULES

S. Posen*[1], G. Wu*[2], E. Harms, A. Grassellino, O. S. Melnychuk, D. A. Sergatskov, N. Solyak
*Fermi National Accelerator Laboratory, Batavia, Illinois, 60510, USA*

A. Palczewski
*Thomas Jefferson National Accelerator Facility, Newport News, VA 23606, USA*

D. Gonnella, T. Peterson
*SLAC National Accelerator Laboratory, Menlo Park, CA 94025, USA*

When a superconducting radiofrequency cavity is cooled through its critical temperature, ambient magnetic flux can become "frozen in" to the superconductor, resulting in degradation of the quality factor. This is especially problematic in applications where quality factor is a cost driver, such as in the CW linac for LCLS-II. Previously, it had been unknown how to prevent flux from being trapped during cooldown in bulk niobium cavities, but recent R&D studies showed near-full flux expulsion can be achieved through high temperature heat treatment and cooling cavities through the superconducting transition with a spatial thermal gradient over the surface. In this paper, we describe the first accelerator implementation of these procedures, in cryomodules that are currently being produced for LCLS-II. We compare the performance of cavities under different conditions of heat treatment and thermal gradient during cooldown, showing a substantial improvement in performance when both are applied, enabling cryomodules to reach and, in many cases, exceed a $Q_0$ of $\sim3\times10^{10}$.

## I. INTRODUCTION

For high energy SRF (superconducting radiofrequency) linacs operating with high duty factor, cryogenic infrastructure and operation can be a major cost driver of the accelerator. With heat load in the low temperature circuit dominated by the dynamic load of the cavity RF heating, the cavity $Q_0$ specification can be key to determining the size of the required cryogenic system. This factor led the LCLS-II project [1] to adopt the "nitrogen-doping" or "N-doping" treatment for the 9-cell niobium cavities in the main linac of their accelerator. Nitrogen doping, developed at Fermilab in 2013 [2], typically allows for $Q_0$ in the range of $3\times10^{10}$ in a 1.3 GHz 9-cell TeSLA-type cavity [3], an improvement of a factor of ~2-3 compared to the previous state-of-the-art at medium accelerating fields ~16 MV/m [2]. However, this $Q_0$ is only realizable in an accelerator cryomodule if extrinsic degradation is avoided. One crucial source of $Q_0$-degradation is magnetic flux "frozen in" during cooldown through the superconducting transition temperature, resulting in heat dissipation when RF fields are applied. In this paper, we start by reviewing the recent developments in basic R&D that suggested a path to improve cavity performance through expulsion of ambient magnetic flux from the cavity during cooldown. We then present measurements of an LCLS-II prototype cryomodule to show what thermal gradients are achievable during cooldown and compare this to measurements of flux expulsion as a function of thermal gradient for several single cell cavities made using the same niobium that would be used in LCLS-II cavity production. We show examples that illustrate the variability observed in flux expulsion behavior for the LCLS-II production material depending on the heat treatment temperature applied to the material. We then give an overview of the performance of LCLS-II cavities measured at Fermilab and show that through proper processing and assembly, it is possible to reliably achieve an average $Q_0$ of $\sim3\times10^{10}$ not just in vertical test but also in cryomodule test. Finally, we discuss the implications for LCLS-II and for future SRF accelerators.

## II. RECENT DEVELOPMENTS IN MAGNETIC FLUX EXPULSION R&D

Magnetic flux expulsion is one of the factors that determine $Q_0$ degradation due to magnetic flux trapped during cooldown. A simple model contains four factors: 1) $B_{amb}$, the ambient magnetic field not generated by thermocurrents, 2) $B_{tc}$, the thermocurrent-generated magnetic field, 3) $\eta$, the fraction of the magnetic field trapped during cooldown (i.e. not expelled), and 4) $S$, the sensitivity to trapped flux, which measures the added surface resistance per unit of flux trapped. The added surface resistance can then be estimated as $S\eta(B_{amb} + B_{tc})$. The ambient magnetic field not generated by thermocurrents is determined by local sources and shields of magnetic flux: the local value of the Earth's magnetic field, the magnetization of the vacuum vessel of the cryomodule, any magnetized components in the cryomodule, the high permeability material used for magnetic shielding in the cryomodule, any active field compensation coils, etc. For an analysis of these items for LCLS-II, see Ref. [4]. Thermocurrent-generated magnetic fields come from thermal gradients across junctions of materials with different Seebeck coefficients, and it may include both static components (e.g. connections on the cavity string to the fixed temperature intercepts) and dynamic components (e.g. temperature differences during cooldown from one cavity-helium vessel weld to the other side). For LCLS-II, the helium vessel, shown in Figure 1, was specially designed to be symmetric to reduce dynamic thermocurrents (note that this design is different from the helium vessels developed for ILC and European XFEL), and studies in the prototype LCLS-II cryomodule found that dynamic thermocurrents were relatively small [5]. The sensitivity to trapped flux determines the amount of $Q_0$ degradation (i.e. increase is surface resistance) for a given quantity of flux trapped in the material. Sensitivity varies depending on the treatment, with a value of ~1.4 nOhm/mG for nitrogen-doped cavities like those in LCLS-II [6].

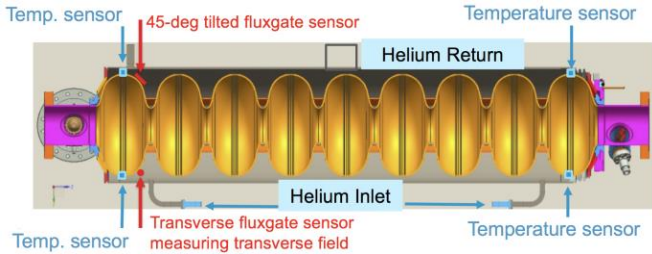

*Figure 1: LCLS-II helium vessel designed to minimize thermocurrents induced by temperature differences between the two ends of the 9-cell cavities, with symmetric helium inlets and central connection to helium return line. For the prototype cryomodule, temperature sensors and fluxgate magnetometers were installed on 4 of the 8 cavities.*

The factors determining the fraction of magnetic flux expelled from bulk niobium cavities during cooldown has been shown to depend on several factors. Studies at Fermilab in 2013 established the importance of cooldown [7], [8]. In these studies, a magnetic field was applied to a single cell 1.3 GHz cavity using Helmholtz coils, and the magnetic field enhancement during cooldown was measured using fluxgate magnetometers. Field enhancement is used as a gauge of flux expulsion – if the cavity fully traps the field, the magnetic field while the cavity is normal conducting $B_{NC}$ is expected to be equal to the field after it becomes superconducting $B_{SC}$; on the other hand, if the field is partially expelled, at the equator of the cavity where the fluxgates are located, there is expected to be some field enhancement after transition: $B_{SC}>B_{NC}$. For a single cell 1.3 GHz cavity, full expulsion of a uniform axial magnetic flux corresponds to $B_{SC}/B_{NC}$~1.7 at the equator [7]. To vary the cooldown conditions, different values of mass flow of cold helium were used during cooldown through the critical temperature, resulting in different spatial thermal gradients across the cavity during transition as measured by temperature sensors on the cavity. These studies established an apparent dependence of the flux expulsion ratio $B_{SC}/B_{NC}$ on the spatial thermal gradient $dT/dx$ during cooldown, with higher $B_{SC}/B_{NC}$ consistently observed for large values of $dT/dx$ (e.g. ~1 K/cm). $Q_0$ was also measured in these cooldown studies, and as expected, the amount of flux expelled during cooldown, as represented by the ratio $B_{SC}/B_{NC}$, correlated strongly with the $Q_0$, and as $B_{SC}/B_{NC}$ approached 1.7, the $Q_0$ approached the value measured when cooling in zero applied field. However, these measurements were made on only one cavity, and it turned out that this cavity had fairly strong expulsion behavior.

Further studies at Fermilab in 2015 [9], [10] showed that different cavities have different expulsion behavior, as measured by $B_{SC}/B_{NC}(dT/dx)$. A thermal gradient that would lead to full expulsion on one cavity may result in only partial expulsion in another cavity, or even near-full trapping. The trend of $B_{SC}/B_{NC}(dT/dx)$ was found to be independent of the surface treatment, but it was found to depend strongly on the production run of the niobium and the heat treatment. A bulk high RRR niobium cavity that was given a standard 800°C 3-hour degassing in a vacuum furnace as part of its treatment may in some cases exhibit poor flux expulsion behavior (i.e. $B_{SC}/B_{NC}$ substantially smaller than 1.7 even at relatively large $dT/dx$~1 K/cm). However, it has been consistently found that flux expulsion behavior of such a cavity can be improved dramatically through the application of heat treatment at temperatures ~900-1000°C for ~3 hours. This was a crucial finding of the R&D program as a way to "cure" cavities with poor flux expulsion behavior.

### III. ACHIEVING HIGH MASS FLOW IN THE LCLS-II MAIN LINAC CRYOMODULES

LCLS-II is a 4 GeV free electron laser under construction at SLAC, for which cryomodules are currently being assembled at Fermilab and Jefferson Lab. The LCLS-II linac will operate in CW (continuous wave) mode. There was an R&D phase in preparation for cryomodule production which included studies of the nitrogen-doping process itself to select the "2/6" implementation [11] as well as industrialization of the treatment at the cavity vendors RI and Zanon [12]. In parallel with the R&D was design of the 1.3 GHz main linac cryomodules at Fermilab, based on the TeSLA cryomodule framework, with modifications for CW operation. A major modification of the design was to allow for high mass flow cooldown in order to improve flux expulsion as indicated by the 2013 Fermilab studies. Rather than cooling the entire linac together, each cryomodule was equipped with its own helium input line, and the 2 K two-phase pipes of the cryomodules were separated from each other, so that a high mass flow from the cryogenic plant could be directed at each cryomodule in turn.

Two prototype cryomodules (pCM) were fabricated, one at Fermilab and another at Jefferson Lab, with specialized instrumentation to measure temperature gradients and thermocurrents during cooldown. $Q_0$ measurements were performed on the Fermilab pCM in four separate cooldowns through the critical temperature, each with a widely different mass flow rate. The highest and lowest mass flow rate cooldowns are shown in Figure 2.

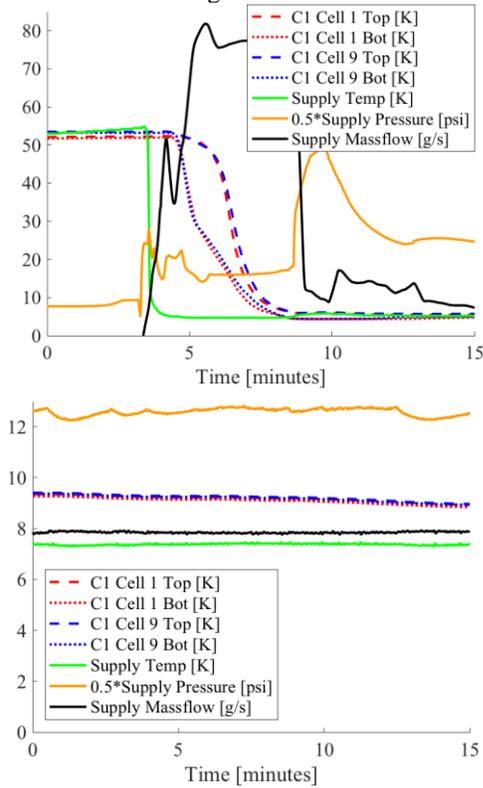

*Figure 2: High (top) and low (bottom) mass flow cooldowns of the Fermilab LCLS-II prototype cryomodule. When 80 g/s mass flow is achieved during cooldown through $T_c$, the temperature difference from the top to the bottom of the cavity is several kelvin, but the difference is much less than 1 K for the 8 g/s cooldown.*

The temperature measurements in Figure 2 show that high mass flow cooldown resulted in successful generation of a significant thermal gradient across the cavity. Meanwhile very low mass flow cooldown resulted in nearly uniform temperature over the surface of the cavity. Figure 3 shows that this was consistently observed across all four of the instrumented cavities in the 8-cavity cryomodule. This figure also shows intermediate mass flow cooldowns, which as expected resulted in intermediate thermal gradients. These measurements in the Fermilab pCM help to establish what thermal gradients are practically achievable in the cryomodule, and therefore what minimum flux expulsion behavior to strive for to minimize $Q_0$ degradation in operation. For additional studies of the Fermilab LCLS-II pCM, see Ref. [5].

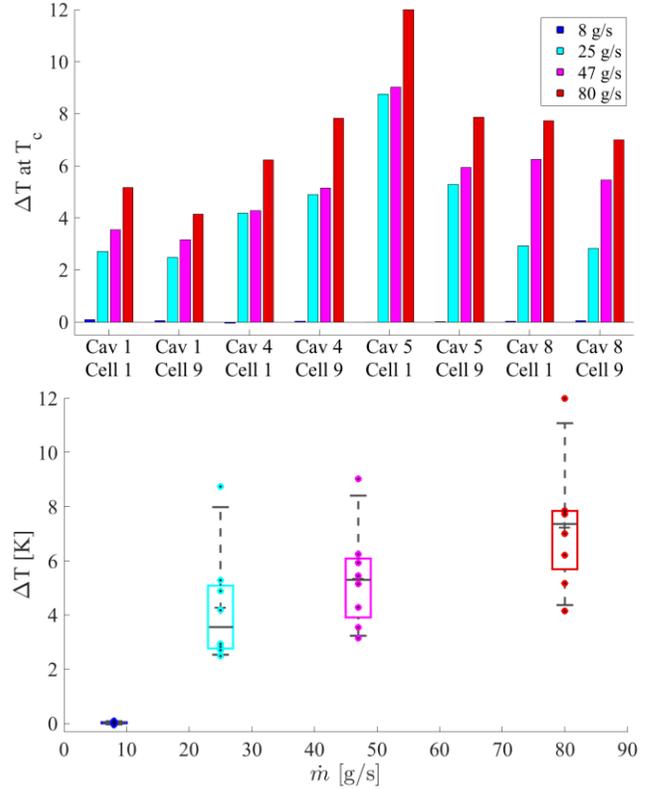

*Figure 3: Top: Temperature difference between the top and the bottom of the cavity achieved during cooldown in cryomodule based on measurements of Fermilab LCLS-II prototype cyromodule. Four different cooldowns were performed, each with a different mass flow (note that mass flow values are measured only for entire cryomodule, not individual cavities). Bottom: The same data plotted as ΔT vs mass flow. 'Box plots' are added to show statistical spread in the data, with dashed lines showing maximum and minimum, colored rectangles showing range of second and third quartile, horizontal black line showing the median, and black '+' showing the mean.*

## IV. MODIFICATION OF LCLS-II CAVITY PROCESSING PARAMETERS TO IMPROVE FLUX EXPULSION

High RRR niobium material for the main linac cavities was procured in 2015 from two companies, Tokyo Denkai (TD) and Ningxia Orient Tantalum (NX). After observing the results of the 2015 Fermilab R&D showing variability in flux expulsion behavior for different production runs of niobium [9], a decision was made to fabricate single cell 1.3 GHz cavities using sample sheets from the high RRR niobium procured for production. Four single cell cavities were made, two from each vendor. Flux expulsion behavior of these cavities is plotted in the bottom part of Figure 4 (the two TD cavities are labelled TD01 and TD02 and the two NX cavities are labelled NX01 and NX02). The top part of the figure shows the statistical spread in cryomodule data from Figure 3, in order to have a comparison to the relevant range of thermal gradient values for different mass flows. The thermal gradient data is generated from differences in temperature measurements over the surface of the cavity and divided by the path length between temperature sensors[1].

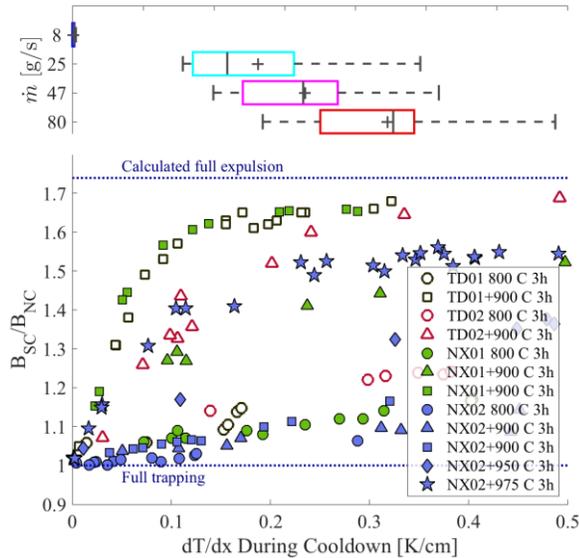

Figure 4: Bottom: Flux expulsion behavior measured by $B_{SC}/B_{NC}(dT/dx)$ in measurements of four single cell 1.3 GHz cavities made from the production material for the LCLS-II main linac. Measurements show that the baseline cavity processing parameters (circles) produce near full trapping behavior for all four cavities. But high temperature treatment of the cavities allows them to achieve near-full expulsion when applying thermal gradients in the range of 0.2-0.3 K/cm during cooldown. Top: Comparison to thermal gradients achieved as a function of mass flow in the prototype cryomodule from Figure 3. The lowest mass flow cooldown of the pCM would be expected to result in nearly full trapping for all cases. The others would result in different levels of flux expulsion depending on the heat treatment of the material.

The four cavities were initially treated using the baseline LCLS-II cavity processing parameters – bulk removal of 140 μm of material by electropolishing (EP)[2], degassing in a vacuum furnace at 800°C for 3 hours followed by 2/6 nitrogen doping, and 5 μm EP. Measurements show near full trapping behavior for all four cavities after this baseline treatment. To improve expulsion, the cavities were reset with light EP (to remove the nitrogen doped material), then given high temperature heat treatment, and nitrogen doped again (doping treatment was applied only after the temperature had stabilized at 800°C). There was variation in how the niobium material responded to the higher temperature heat treatment. Of the two single cell cavities made with TD material, after heat treatment at 900°C for 3 hours, one showed very strong flux expulsion behavior and the other was fairly strong. Of the two NX cavities, one showed strong flux expulsion after an accumulated 6 hours at 900°C, and the other only achieved reasonably strong flux expulsion behavior after four 3-hour furnace runs at increasing temperatures up to 975 C.

In response to the single cell measurements, it was decided to heat treat all of the TD material at 900°C for 3 hours. In the early part of cavity production, the same temperature was applied to the NX material, but after observing initial results, an increased temperature of 950°C for 3 hours was later selected for the "lot A" and "lot B" NX material and 975°C for 3 hours for the "lot C" NX material (like that in single cell cavity TD02). Heat treatment temperatures were chosen conservatively low, to avoid compromising the mechanical properties of the cavities by having too low yield strength[3]. The high temperature heat treatment would be applied in production during the furnace run that occurs as part of the cavity processing sequence (i.e. not on the raw material, but by elevating the temperature of the degassing step at prior to 2/6 doping at 800°C). In addition, the amount of material removed during bulk EP was increased from 140 μm to 200 μm to try to more fully remove the so-called "damage-layer" created by processing (e.g. rolling) that may contribute to residual resistance. The plan to modify the processing parameters was put into action after the first sequence of 16 cavities had been produced (serial numbers CAV0001-CAV0016). CAV0017 was the first cavity with higher temperature heat treatment and deeper removal. It was sent to Fermilab without a vacuum vessel so that detailed measurements could be performed. Figure 5 shows a cooldown of this cavity which was heavily instrumented with 8 temperature sensors (one on every cell except cell 5) and 4 fluxgate magnetometers (on cells 2, 4, 6, and 8). Note that as each of the cells with fluxgate magnetometers goes through the critical temperature of 9.2 K, a strong field enhancement is observed in the magnetic field probe corresponding to that cell. The measured ratio $B_{SC}/B_{NC}$ in each case is fairly close to what is calculated for full expulsion in a 9-cell cavity, ~1.4.

---

[1] Iris-to-iris distance for the single cell vertical test data, 187 mm; half the outer equatorial circumference for the cryomodule data, 324 mm.
[2] Material removal with EP is generally non-uniform, with more removal at the irises and less at the equators. In this paper, we use the convention of defining material removal averaged over the interior surface of the cavity.
[3] Frequency shift during shipment of dressed cavities from the vendor can be an indication of excessive reduction of structural strength. However even after higher temperature heat treatment, measurements were consistently well within the acceptable range.

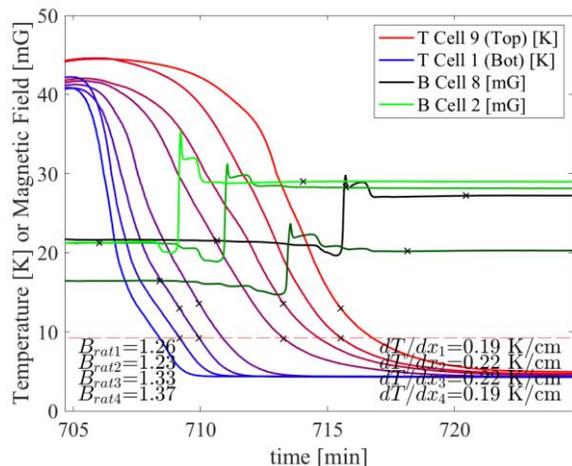

*Figure 5: Example of 9-cell cooldown of cavity in VTS without helium jacket. The lines with colors ranging from blue to red show the temperature data for cells 1 (bottom of cavity), 2, 3, 4, 6, 7, 8, and 9 (top of cavity). The lines with colors ranging from green to black show the magnetic field data for cells 2, 4, 6, and 8. The "jumps" in the magnetic field data line up well with the expected flux expulsion signal occurring as the respective cells pass through the critical temperature 9.2 K. $B_{rat}$ is used here to denote the ratio $B_{SC}/B_{NC}$ for each of the four fluxgate magnetometers.*

CAV0019 was the first dressed cavity with higher temperature heat treatment and deeper removal. In Figure 6, its performance is compared to that of CAV0007, which was in the first production sequence, before the modifications to the cavity processing parameters. Each of these cavities was tested in three different ambient magnetic field conditions: <1 mG (i.e. compensated field using feedback from fluxgate to coil), ~5 mG, and ~10 mG. CAV007 shows strong degradation when a field is applied, below $2 \times 10^{10}$ in 5 mG. This is noteworthy, as 5 mG is the specification for the ambient magnetic field at the cavity in the LCLS-II cryomodule, and exposing CAV0007 to this level of applied field degrades the $Q_0$ to significantly below the $Q_0$ specification. However, CAV0019 shows very little $Q_0$ degradation even for 10 mG ambient field, still showing $Q_0$ well above specification. The difference between the two cavity $Q_0$ values after cooling in <1 mG field may be due to differences in expulsion of the small residual fields, or possibly due to reduced residual resistance from additional removal in the bulk EP step. Note that for this vertical test data and for all 9-cell vertical test data presented in this paper, the cavities were tested with stainless steel flanges connected to their beam ports, which adds approximately 0.8 nΩ to the overall surface resistance. The $Q_0$ specification is lowered from $2.7 \times 10^{10}$ to $2.5 \times 10^{10}$ for vertical testing to account for this.

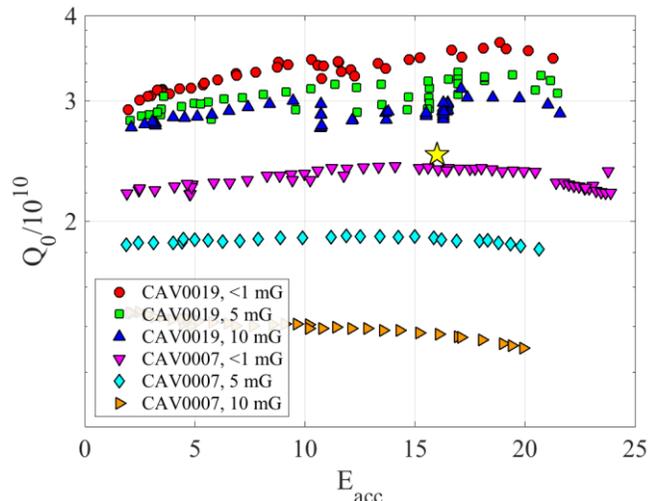

*Figure 6: Comparison of 2.0 K Q vs E vertical test data for CAV0007, which received the baseline cavity processing parameters, including 140 μm bulk removal via electropolishing and 800°C 3 hours heat treatment, to CAV0019, which received the modified treatment of 200 μm bulk removal via electropolishing and 900°C 3 hours heat treatment. Both are made with TD material and both cavities were dressed with helium jacket. Measurements were made for cooldowns in various values of external field. The stainless steel flanges add approximately 0.8 nΩ to the overall surface resistance.*

Fermilab's pCM, while built as a prototype, is also planned to be used in the linac, and it has the serial number CM01 (full serial number F1.3-01 to denote its frequency and the lab where it was assembled—in this paper we condense to simply CM01 as we are examining only Fermilab 1.3 GHz cryomodules). CM01 contains cavities that were fabricated several years ago under a different program, but were repurposed, including removal of the ILC-style helium vessel, application of the nitrogen-doping treatment, and welding of the LCLS-II vacuum vessel. CM01 cavities are made from ATI Wah Chang (WC) material. CM02 cavities are from TD material using the baseline treatment parameters including 800°C heat treatment and 140 μm removal (this includes CAV0007). CM03 cavities are from TD material and were the first that were treated at 900°C with 200 μm of removal (similar to CAV0019). The difference in the performance of these two sets of cavities is clearly seen in the vertical test data in Figure 7. There is a wide separation between the CM02 cavities—the majority of which are just below the $Q_0$ specification—and the CM03 data—which are all well above specification. At the time of testing, there was a 24 MV/m administrative limit on the accelerating gradient in vertical test, though this restriction was later removed. As before, stainless steel flanges add approximately 0.8 nΩ to the surface resistance. CAV0032 shows a $Q_0$ degradation after quenching at ~12 MV/m, likely due to flux trapping. Also plotted on the graph are comparison points to measurements of cavities treated with 120°C bake (i.e. not nitrogen treated) from the European XFEL, based on data from [13].

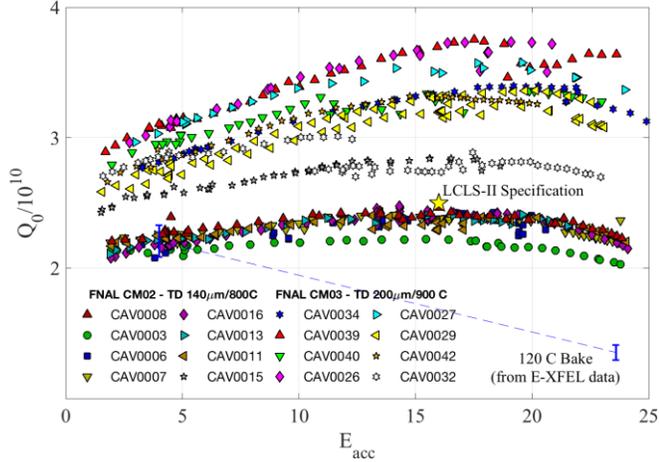

*Figure 7: Vertical test data for cavities from CM02 and CM03. Cavities from CM02 had 800°C 3-hour heat treatment and bulk removal of just 140 μm while cavities from CM03 had 900°C 3-hour heat treatment and bulk removal of 200 μm, which results in a consistently higher Q0. Data are measured at 2.0 K after fast cooldown.*

At the time of writing of this paper, 13 cryomodules have been gone through qualification testing at Fermilab's cryomodule test facility (CMTF), each of which had their cavities qualified in Fermilab's vertical test stand (VTS) prior to cryomodule assembly. Figure 8 shows $Q_0$ measurements for all cavities to go through both types of testing so far. Measurements were made at a temperature of 2.0 K and an accelerating gradient of 16 MV/m (the operating gradient planned for LCLS-II—the small fraction of cavities that did not reach this field in cryomodule test are not shown in this plot). For VTS, cavities were generally cooled through the critical temperature with high mass flow so large thermal gradients are expected for each test. For CMTF, some cryomodules have $Q_0$ data measured for multiple cooldowns; in these cases, the data is presented for the cooldown with the highest mass flow (at least 30 g/s for all CMTF data in this plot). Note that there is some difficulty in comparing VTS data, as early measurements were performed in compensated magnetic field, but as production progressed, testing was carried out without field coils. Also, it should be noted that there is some variation in residual magnetic field in the three test dewars at Fermilab (though all are expected to give at most ~5 mG ambient field at $T_c$). These variations are not considered in the plot, and they are not anticipated to have a significant effect on the analysis and conclusions presented here.

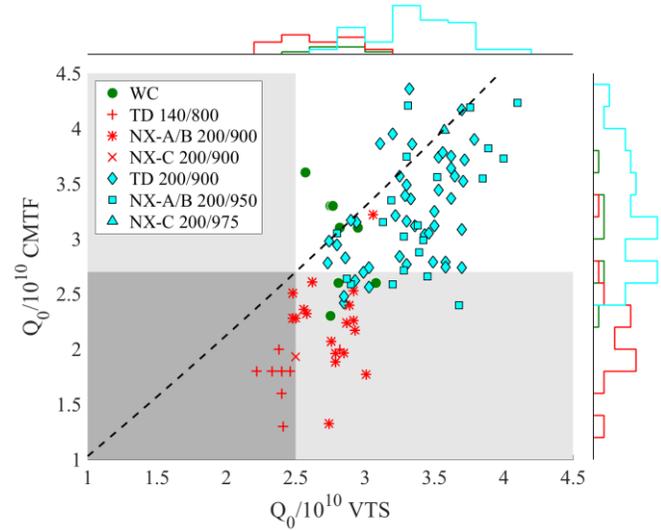

*Figure 8: Q0 measurements of each cavity in Fermilab LCLS-II CM01-13, showing a strong correlation to the heat treatment temperature of the cavity for a given material. The red symbols +, ✱, and × represent cavities that were treated at too low of a temperature for that material to have strong flux expulsion (see CAV0007 in Figure 6). The blue symbols ◇, ▢, and △ represent cavities that were heated at a sufficiently high temperature (see CAV0019 in Figure 6). The green symbols ● represent cavities from the prototype cryomodule, and are considered separately, as they were fabricated previously and therefore not using LCLS-II production material. Q0 data are shown for measurements in both vertical test (VTS) and cryomodule test (CMTF). The gray boxes are related to the Q0 specifications as explained in the text. Histograms on the top and right side of the scatter plot show counts of cavities binned by Q0, separated into three categories corresponding to the color of the symbols in the scatter plot. All data measured at 16 MV/m and 2.0 K after high mass flow cooldown. Note that stainless steel flanges are expected to add ~0.8 nΩ of surface resistance, which is accounted for by the dashed line.*

The legend explains how the material lot and heat treatment are represented for each of the cavities. For early production cavities, represented by the red symbols, heat treatment was insufficient to achieve strong flux expulsion, as demonstrated in Figure 6. This includes TD material heat treated at 800°C for 3 hours and NX material at 900°C for 3 hours. As indicated by measurements in Figure 4, these cavities are expected to strongly trap ambient fields even for a high mass flow cooldown. This is borne out in both the VTS and CTMF data. Cavities fabricated later in production received heat treatment at sufficiently high temperature to result in strong expulsion. These cavities, represented by blue symbols, include TD material heat treated at 900°C for 3 hours, NX lot A and lot B material at 950°C for 3 hours, and NX lot C material at 975°C for 3 hours. These cavities consistently show higher $Q_0$, both in VTS and CMTF. The pCM cavities made from WC material are included as well, though the expulsion behavior of this material is less known. The material removal and heat treatment for the pCM cavities is not shown as it varies depending on the history of the cavities.

The regions of the plot below the nominal specifications in VTS and CMTF are indicated by the gray boxes. However, the CMTF "specification" of $2.7\times10^{10}$ is more of a nominal target for $Q_0$ chosen according to the average dynamic load anticipated per cavity to fit within the limits of the cryogenic plant. Lower $Q_0$ values can be accepted for some cavities depending on the $Q_0$ of other cavities and the balancing of accelerating gradients when the linac is in operation. This balancing is considered in more detail in Section V.

The dashed line indicates where cavities should lie if they have equal $Q_0$ in VTS and CMTF, after accounting for the stainless steel flange losses in VTS. The fact that the vast majority of cavities fall below the curve indicates a general trend of degradation of $Q_0$ at CMTF compared to VTS. One cause for this may be that generally when cooling from room temperature in the dewar, higher thermal gradient could be achieved than those observed in the cryomodule even with 80 g/s mass flow. Another cause for lower $Q_0$ in the cryomodule might be higher ambient fields, both thermocurrent generated and non-thermocurrent generated.

Note that the material designation (e.g. "NX-A/B" or "NX-C") refers to the dominant material type in a cavity. Before the flux expulsion behavior was well understood, a number of cavities were fabricated with a mixture of NX material from lots A, B, and C. In addition, two cavities were fabricated using predominantly TD material but with one endgroup made with NX material.

Figure 10 presents the Fermilab CMTF data from Figure 8 separated by cryomodule, from CM01 through CM13. The cavity material and heat treatment are again indicated on the plot, using the same color scheme as Figure 8 to indicate production cavities that are expected to have sufficient/insufficient heat treatment to cause strong flux expulsion. Within each cluster representing a cryomodule, the cavities are ordered by their position in the cryomodule, ending with the cavity adjacent to the quadrupole magnet. Accelerating gradient measurements are shown below the $Q_0$ data, including the onset of field emission, the point at which the field emission exceeds 50 mR/hr (set as an administrative limit for "usable" gradient), and the approximate onset of sporadic quench behavior, in which the cavity may operate at a given gradient for several minutes or tens of minutes then quench suddenly. In most cases, sporadic quenching is expected to be caused by multipacting—due to limited CMTF testing time, only a small amount of processing was performed, and additional time spent processing during linac commissioning is expected to increase the maximum usable gradient of these cavities.

\

The CMTF $Q_0$ data in Figure 10 are nearly the same as those in Figure 8, except in this case the small fraction of cavities that do not reach 16 MV/m are also shown, using the $Q_0$ measurement at the highest "usable" gradient. The correlation between $Q_0$ and heat treatment (represented by the blue or red colors) is again evident, but additional information is gained by clustering the data by cryomodule. Modules 5, 6, 7, and 8 are especially illuminating, as these modules each contain both cavities that have had sufficient heat treatment for strong expulsion (TD after 900°C 3 hours or NX lot "A" and "B" after 950°C 3 hours) and cavities with heat treatment below the threshold for strong expulsion (NX lots "A," "B," and "C" after 900°C 3 hours). These cryomodules illustrate the effect of flux expulsion on $Q_0$ clearly. All 8 cavities should see very similar thermal gradient and magnetic field during cooldown, but there is a readily apparent trend of substantially lower $Q_0$ from the cavities with insufficient heat treatment.

It should be noted that particularly for some of the earlier cryomodules, the cooldown procedures were still being developed, which is expected to contribute to not fully optimized performance. For example, staring with CM07, the procedure was modified so that all coupler intercepts and shields were kept as cold as possible during cooldown.

Figure 9 illustrates how mass flow during cryomodule cooldown affects $Q_0$. For a given cryomodule cooldown, all cavities are averaged together[4]. The expected trend is observed: higher mass flow leads to higher $Q_0$. The only exception is CM02, possibly due to imperfect cooldown procedures, incomplete thermalization in the first cooldown, or dynamic thermocurrents.

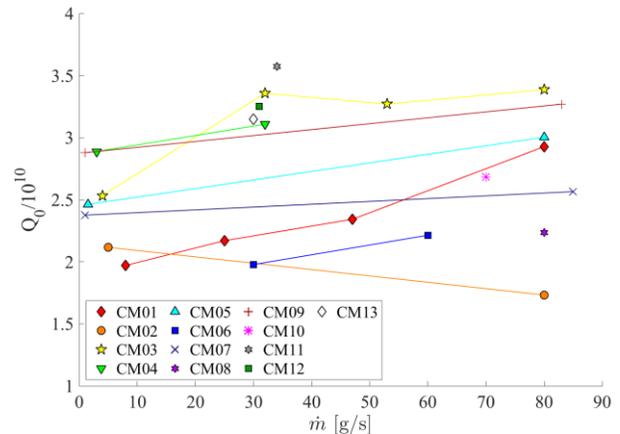

*Figure 9: Cryomodule Q0 as a function of mass flow of cold helium during cooldown through the critical temperature. In nearly all cases, higher Q0 is achieved at higher mass flow.*

---

[4] Averaging is done at the heatload level, so in this case the $Q_0$ plotted in Figure 9 is given by the harmonic mean (i.e. the reciprocal of the average of the reciprocals) of the $Q_0$ values of the eight cavities in the cryomodule.

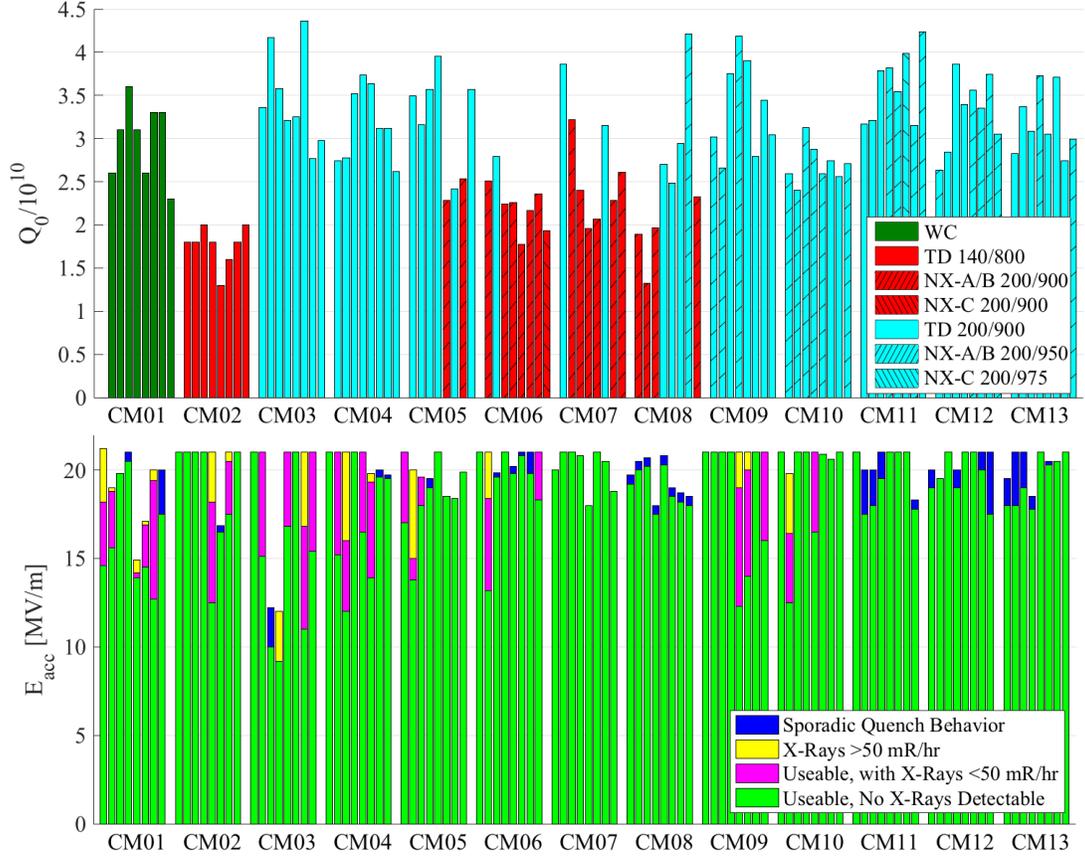

*Figure 10: Top: $Q_0$ measured in CMTF as a function of cryomodule number and position in cryomodule. Q0 correlates strongly with whether the heat treatment is expected to be sufficient (blue) or insufficient (red) to cause strong flux expulsion. The prototype cryomodule CM01 was made with non-production material (green). All data measured at 16 MV/m and 2.0 K after high mass flow cooldown. Hatching indicates material lot. Bottom: $E_{acc}$ measurements of each cavity in the cryomodule. Additional details are given in the text regarding the color scheme.*

## V. DISCUSSION

A variety of results were presented in section IV. This includes $Q_0$ measurements from vertical test in Figure 6, Figure 7, and Figure 8 and from cryomodule test in Figure 8, Figure 10, and Figure 9. The measurements were presented with LCLS-II production cavities separated into two broad categories: those that were heat treated at sufficiently high temperature to result in strong flux expulsion behavior, and those that were not. The substantially higher performance of cavities in the first category is striking. For example, for the data in Figure 8, the centroid of the cavities that were sufficiently heat treated for strong expulsion is $Q_{0,VTS} = (3.4 \pm 0.3) \times 10^{10}$, $Q_{0,CMTF} = (3.3 \pm 0.5) \times 10^{10}$, and the centroid for those that were not is $Q_{0,VTS} = (2.7 \pm 0.2) \times 10^{10}$, $Q_{0,CMTF} = (2.1 \pm 0.4) \times 10^{10}$. While there is a substantial difference depending on the heat treatment, it is important to consider that a $Q_0$ in production above $2 \times 10^{10}$ at 16 MV/m and 2 K in a TeSLA type cavity would have been well beyond the expected performance with previous state-of-the-art cavity processing techniques such as EP/120°C bake. In other words, even without optimized heat treatment, N-doping enables $Q_0$ far exceeding what was possible previously, but realizing the full potential of doped cavities requires trapped flux to be minimized.

The cavities that fall into the category of having weak flux expulsion behavior due to insufficient heat treatment will be vulnerable to any residual fields during cooldown. For these cavities, the precautions taken to lower ambient fields in the cryomodule will be especially important. This includes careful magnetic hygiene, minimization of thermocurrents, and demagnetization of the vacuum vessel [4]. If nearly all ambient flux is trapped in these cavities, these precautions will help to minimize the degradation.

There was only one category of material with bulk material removal less than 200 μm: TD material with 140 μm bulk removal and 800°C 3 hours heat treatment (TD140/800). This material also shows some of the lowest $Q_0$ values in both vertical and cryomodule test. While weak flux expulsion is expected to play a strong role, increased residual resistance due to insufficient removal may be causing further degradation for these cavities.

For the particular production lots of niobium used to fabricate LCLS-II cavities, the NX material required higher temperature heat treatment to achieve strong expulsion behavior than the TD material. However, once the optimized heat treatment was applied, material from both vendors was made to exceed the requirements of the project. And further, for the $Q_0$ data in Figure 8 from cavities that were heat treated at sufficiently high temperature (blue points), there is no obvious clustering by vendor. Once sufficiently heat treated, niobium from both vendors was equivalently qualified to produce cavities with $Q_0 \sim 3\times10^{10}$ or higher.

Figure 10 shows that there are a number of cavities with $Q_0$ exceeding the nominal goal of $2.7\times10^{10}$ and that also have a maximum usable gradient above the nominal 16 MV/m. While operating these cavities with high $Q_0$ at 16 MV/m leaves margin for cavities with poorer $Q_0$, they can also be operated at higher gradient to allow the gradient to be turned reduced on cavities with lower $Q_0$. This optimization process should be performed on the linac as a whole, but to illustrate the idea, an example is shown considering just the cavities in CM07 in Table 1. The dynamic load is calculated for the measured $Q_0$ with all the cavities operating at 16 MV/m. In the next column, the gradients are modified up or down by 2 MV/m and the dynamic load is recalculated under the assumption that the $Q_0$ is relatively constant in the optimization range (a fairly reasonable assumption based on the curves in Figure 7). The dynamic load is reduced compared to the nominal gradients. In the last column, gradients are presented from a simple optimization algorithm that results in a more than 10% reduction in dynamic load for these 8 cavities. The optimization is limited by the maximum usable gradient for each cavity, and in fact the optimization puts Cavity 1 at its limit. For each of the three different cases, the overall accelerating voltage of the cavities taken together remains the same.

*Table 1: Example to illustrate cavity-by-cavity adjustment of the gradient to minimize dynamic load.*

| Cavity | $Q_0$ /$10^{10}$ | Nominal [MV/m] | Modified [MV/m] | Optimized [MV/m] |
|---|---|---|---|---|
| 1 | 3.86 | 16 | 18 | 20.0 |
| 2 | 3.22 | 16 | 18 | 18.8 |
| 3 | 2.40 | 16 | 14 | 14.0 |
| 4 | 1.96 | 16 | 14 | 11.4 |
| 5 | 2.07 | 16 | 14 | 12.1 |
| 6 | 3.15 | 16 | 18 | 18.4 |
| 7 | 2.28 | 16 | 14 | 13.3 |
| 8 | 2.61 | 16 | 18 | 15.2 |
| **Dyn. Load [W]** | | **83.0** | **80.5** | **73.5** |

Figure 9 shows the importance of achieving high mass flow during cooldown to minimize degradation of $Q_0$ due to flux trapping. Unfortunately, the statistics are limited in most cryomodules due to the financial and time constraints of testing during production. Furthermore, the Fermilab dataset is the primary source of this data as the Jefferson Lab cryomodule testing facility was only able to produce fairly low mass flow cooldowns (<8 g/s) until it was upgraded around the testing of Jefferson Lab's CM08. However, there are two Fermilab cryomodules in which 4 cooldowns were performed with a variety of mass flows: CM01 and CM03. For CM03, approximately the same $Q_0$ was obtained for all 3 cooldowns with mass flow ~30 g/s or higher. It is useful to compare this to the data presented in Figure 4. Measurements of single cell cavities made from LCLS-II production material show that when sufficiently high heat treatment is applied resulting in strong flux expulsion behavior, near-complete expulsion is achieved for cooldown with any thermal gradient higher than ~0.2 K/cm. The figure also shows that this gradient corresponds to a cryomodule mass flow of ~30 g/s. The saturation of CM03 $Q_0$ in Figure 9 is therefore consistent with achieving near-complete flux expulsion. For CM01 on the other hand, improvement in $Q_0$ was observed consistently from 8 g/s up to 80 g/s cooldown, consistent with material showing weaker flux expulsion behavior.

In the SLAC tunnel, where the cryomodules will be placed, fast cool downs will also be possible. The 4 kW cryoplants are capable of a helium gas flow of ~120 g/s. While in principle this large amount of mass flow can be used to cool individual cryomodules, the practicalities of doing so make it difficult. The linac is going to be divided into 4 sections: L0 comprised of one cryomodule, L1 comprised of two 3.9 GHz and two 1.3 GHz cryomodules, L2 comprised of 12 cryomodules, and L3 comprised of 20 cryomodules. The tunnel itself is sloped downward from L0 at a 0.5% grade. The helium distribution enters the tunnel between L2 and L3, with one distribution line pushing helium uphill and one downhill. This slope and difference in helium distribution makes the cooling of the different sections non-identical.

During operation, cryomodules will be cooled from ~40 K to 4.2 K one at a time, to maximize the cooling on each. However, the exact speed of gas flow must be limited depending on which section of the linac is being cooled. In order to understand the mechanics of helium flow in the linac for the different sections, a study was commissioned by LCLS-II through the Technische Universität Dresden [14]. This study modeled helium flow uphill versus downhill in the linac configuration at SLAC. The main concern is that due to the series connection of each cryomodule to the next in a given linac section, helium from one cryomodule being cooled could spill over into the next through the helium gas return pipe. This could cause an inadvertent slow cool down on adjacent cryomodules, reducing the flux expulsion efficiency. For cryomodules in L0, L1, and L2, helium is pushed uphill, resulting in significant mitigation of this effect if cryomodules are cooled starting with the most-downstream cryomodule and working uphill. Helium gas flow rates of up to 80 g/s were modeled in this region and it was found that cryomodules could be cooled sequentially without inadvertent slow cool down of the adjacent cryomodule. In L3, where helium flows downhill, this situation is different. With very fast helium flow rates, such as 80 g/s, inadvertent slow cool down of adjacent cryomodules would occur due to helium spillover. However, with moderately fast cool downs using 30 g/s of helium flow or less, this result is mitigated. Therefore in L0, L1, and L2, cryomodules can be cooled with helium flow rates upwards of 80 g/s, but in L3, cryomodules can only be cooled with flow rates of 30 g/s at the highest.

The limitations on the cooling capabilities due to the tunnel slope create better and worse locations for certain cryomodules. Ideally, cryomodules comprised primarily of cavities with poor flux expelling material should be placed in L0, L1, or L2, where they can be cooled sufficiently fast. For cryomodules comprised of good flux expelling material, any location in the SLAC tunnel will suffice. In reality, LCLS-II will not specifically sort cryomodules based on material. However, since poor flux expelling cavities are primarily located in early production cryomodules, most cryomodules that should be placed in L0, L1, and L2 will be.

The study in this paper focuses on flux expulsion in N-doped cavities for LCLS-II, but the implications are important for all bulk niobium SRF cavities. For example, for a linac with 120°C baked cavities (e.g. European XFEL, ILC, ESS etc.), the sensitivity to trapped flux at 16 MV/m and 1.3 GHz will be lower (approximately 0.5 nΩ /mG) than it would be for N-doped cavities (approximately 1.4 nΩ/mG), but the sensitivity increases substantially with gradient (approximately 1.4 nΩ/mG at 35 MV/m) [15]. In these cases, it may be worthwhile to increase heat treatment temperature to improve flux expulsion and in this way reduce the cryogenic component of the overall cost (even if it is not as much of a cost driver as it is in LCLS-II). It would therefore be beneficial for future SRF accelerator projects to improve understanding of the physics of flux expulsion and develop methods to reliably predict without extensive cavity experiments what heat treatment temperature would be sufficient to achieve strong expulsion behavior for a given production lot (without using overly high temperatures that may compromise mechanical properties). Improved understanding could lead to improved niobium material specifications for predictable flux expulsion.

## VI. CONCLUSIONS

SRF cavities made with bulk niobium can be vulnerable to $Q_0$ degradation due to flux trapped in the superconductor during cooldown. In this paper, we showed that by applying lessons learned from recent R&D studies on flux expulsion—specifically applying a large thermal gradient during cooldown and properly heat-treating cavities—it was possible to mitigate extrinsic $Q_0$ degradation due to trapped flux in cryomodules produced for LCLS-II and achieve $Q_0$ close to the intrinsic performance of the N-doped cavities. We evaluated the performance of LCLS-II production cavities measured at Fermilab to date, showing that cavities with sufficient heat treatment for a given material achieve on average a $Q_0$ 26% higher in vertical test and 56% higher in cryomodule test compared to those with insufficient heat treatment. We showed that with the combination of N-doping and strong flux expulsion, LCLS-II cryomodules are being produced at Fermilab that consistently reach and, in many cases, exceed a $Q_0$ of ~3×10$^{10}$ under the operating conditions of LCLS-II. The highest performing cavities will be used to balance those with weaker flux expulsion or other degradation through an optimization of gradient and placement in the linac. Continued development of fundamental understanding of flux expulsion and continued development of material specifications to improve flux expulsion will help to achieve the highest $Q_0$ in future bulk niobium SRF applications.

## ACKNOWLEDGEMENTS

The authors would like to thank the many people who contributed to the results presented in this paper. This includes the Fermilab VTS team, the teams responsible for designing, assembling, and testing the cryomodules at Fermilab, and the LCLS-II project teams at Fermilab, Jefferson Lab, and SLAC. This work was supported by the United States Department of Energy, Offices of High Energy Physics and Basic Energy Sciences under Contracts DE-AC05-06OR23177 (Fermilab), DE-AC05-06OR23177 (Jefferson Lab), and DE-AC02-76F00515 (SLAC).